\documentclass[aps,preprint]{revtex4}
\usepackage{epsfig}
\usepackage{pstricks}
\usepackage{amsmath}
\begin{document}

\title{Dynamic Matter-Wave Pulse Shaping}

\author{M. Nest$^1$, Y. Japha$^2$, R. Folman$^2$, R. Kosloff$^3$}

\affiliation{$^1$Technische Universit\"at M\"unchen, Lichtenbergstr. 4, 85747 M\"unchen,
Germany \\
$^2$Ben-Gurion University of the Negev, Department of Physics, Beer Sheva
84105, Israel \\
$^3$Hebrew University, Department of Physical Chemistry, Jerusalem 91904,
Israel}

\begin{abstract}
In this paper we discuss possibilities to manipulate a matter-wave with time-dependent
potentials. Assuming a specific setup on an atom chip,
we explore how one can focus, accelerate, reflect, and stop an
atomic wave packet, with, for example, electric fields from an array of
electrodes. We also utilize this method to initiate coherent splitting or an arbitrary 
wave form. Special emphasis is put on the robustness of the control schemes.
We begin with the wave packet of a single atom, and extend this to a BEC, in the
Gross-Pitaevskii picture. In analogy to laser pulse shaping with its wide variety of
applications, we expect this work to form the base for more complex time-dependent
potentials eventually leading to matter-wave pulse shaping with numerous applications.
\end{abstract}

\maketitle

\section{Introduction}
Light pulse shaping has given rise to numerous versatile applications
\cite{gerber,woeste08,woeste07,hertel,malinovskaya,friedland}. In
the past decade the control of matter-waves has become comparable in its
accuracy to the control over light. Production, manipulation and measurement of
cold atoms with a de-Broglie wave length of the order of a micron has become
common practice. It is therefore interesting to see whether matter-wave pulse-shaping,
analogous in some way to the optical counterpart, is possible, and what the
similarities and differences are. While we do not claim to have realized in this work
a complete set of universal operations enabling any matter-wave pulse shaping target,
we have indeed demonstrated several fundamental operations for the manipulation of
matter-waves.

As a specific example for the analogous role the two pulse shaping realms may
play, one can study the topic of control of chemical reactions. Such control is
one of the highest goals in today's theoretical chemistry. Over the last ten
years, the concept of coherent control has been developed, as one of the ways to
achieve the desired control. This concept has up to now mostly been used based
on coherent laser light pulses, that were optimized to perform a certain task,
like steering a chemical process towards a specific reaction channel. The
flexibility of these control schemes, like Optimal Control
Theory (OCT) \cite{rabitz88,ronnie89,brumer98,fujimura98,kroener00,klamroth06,
saalfrank06} or some stochastic schemes \cite{rabitz03,woeste06,klamroth08},
stems from the fact that laser pulses can be shaped in many different ways, by 
employing chirping, pulse trains, etc. Recently a different approach to 
coherent control has emerged, which relies on the coherent properties of 
matter-waves. In an example exhibiting the potential of such a method, 
J\o{}rgensen and Kosloff \cite{kosloff2pacc} used the parameters of two 
matter wave packets to control an Eley-Rideal reaction.

To examine if a realistic scheme for matter-wave pulse shaping is feasible, two
things have to be considered. The first is the natural dispersion of
matter-waves, and the second are the experimental constraints. The effect of
dispersion can best be illustrated by comparison with the free propagation of a
chirped laser pulse. Consider a light pulse with a chirp, which is red at the
beginning and blue at its end. This order will be preserved as long as the pulse
travels through vacuum. The atom laser analog would be a wave packet with a
small momentum at the front, and a high momentum at its trailing part.
A configuration which corresponds to this momentum chirp may be naturally
achieved for atoms by using a parabolic potential which accelerates the rear part
relative to the front part of the wave-packet. However, the consequent dynamics is
different than that of light in vacuum. The trailing part will overtake the slower 
frontal part, reversing the order, with possibly some complex interference pattern 
in the process. (This is not quite true anymore when more than one particle is 
considered, see Section III.) Thus, there are limitations on the realization of chirping 
for matter-waves, which will also apply to other control tasks.

Second, one has to take into account the basic experimental setting. In this
paper we take as an example the atom chip \cite{reichel02,henkel02,zimmermann07},
which provides a unique micro-lab for experimentation with ultra cold gases and
Bose-Einstein Condensates (BEC) \cite{reichel2001,zimmermann2001}. The chip has
micro and nano-structured wires and electrodes on its surface \cite{plamen09,ran09},
creating static magnetic or electric fields, as well as micro-wave and radio-frequency
fields to trap, guide and manipulate clouds of neutral atoms as close as a few hundred
nano-meters from the surface (below that the Casimir-Polder force overcomes the other
forces). The atom chip has proven to enable spatial coherence close to the surface
\cite{schmiedmayer05}.

Previous schemes for the manipulation of atomic wave-packets were
based mainly on harmonic potential traps whose harmonic parameters,
namely the frequency and center, may be changed in time to control either
the width of the wave-packet \cite{Castin96} or its center-of-mass
position \cite{Japha02,Ott03}. Other schemes are based on periodic light potentials 
\cite{Oberthaler03, Oberthaler04, Cornell05}. All these schemes maintain potential 
symmetries which limit the possibility to create arbitrary wave forms. To make a step 
forward, here we wish to use optimization methods and arbitrary potentials towards a 
general scheme for matter-wave pulse shaping.

In this paper we study the possibility to manipulate the shape of a coherent
wave packet with an array of electrodes that produce a quasi-static electric
field. Such an array, combined with a magnetic guide to create a potential
barrier between the atoms and the surface, has been successfully employed
\cite{schmiedmayer03}. The setup is such, that the wave packet travels along a
one-dimensional (1D) magnetic wave guide over the array of electrodes. In the
quasi-static regime, the interaction between wave packet and electric field is
given by the potential energy
\begin{equation}
V(x,t) = -\frac{\alpha}{2} |E(x,t)|^2 \quad ,
\label{elecpot}
\end{equation}
where $x$ is the coordinate along the wave guide, $\alpha$ is the static
polarizability, and $E$ is the field generated by the electrodes. The electrodes
can be switched with a time-scale of micro-seconds and below, and may have a distance
of less than a micrometer from each other. The fabrication limitation and atom-surface
distance noted above limit the spatial resolution in constructing such a potential to
a few hundred nanometers, which is much smaller than the wave packet extent and is
sufficient for pulse shaping needs. In this work we utilize temporal and spatial
resolutions which are much less demanding than those noted above.

Given this basic setup, we will explore in the following sections the possibility to
shape an atom laser pulse or simply a wave packet released from its trap into
the guide. As fundamental building blocks of matter-wave shaping, several basic
shaping/control tasks are considered in the following: focusing, accelerating,
reflecting, and stopping the wave packet. Finally, we show how the same technique
could be used for the more complex tasks of coherently splitting the wave packet, and 
creating an arbitrary wave form.

For each task, we discuss how it can be achieved, how sensitive the result depends on 
the parameters of the external potential, and what differences are to be expected
if a BEC, described by the Gross-Pitaevskii equation
\cite{leggettbec,castin98,gardiner97}, is considered. Namely, we put special
emphasis on the robustness of the control schemes. Similar control tasks have
been studied before \cite{esslinger01}, but by means of optical transitions. Let us note 
that optical schemes such as the latter or the previously mentioned optical manipulation 
techniques \cite{Oberthaler03, Oberthaler04, Cornell05}, have a limited resolution 
(relative to the de-Broglie wave length), and are hard to perform close to material 
objects, such as those required in chemical reactions as noted in the above example 
\cite{kosloff2pacc}.

In section II we give a brief description of the theoretical methods used
in this paper, section III analyzes the results of the basic tasks described above, 
while section IV describes the more complex tasks of coherent splitting, as well as the 
creation of an arbitrary wave form. Section V summarizes and gives an outlook.

\section{Theory}

We begin by
solving the time-dependent Schr\"odinger equation (TD-SE)
\begin{equation}
\dot{\Psi} = -\frac{i}{\hbar} H \Psi
\label{tdse}
\end{equation}
for a single $^{87}$Rb atom on the atom chip. Here the Hamiltonian $H$ is given
by
\begin{equation}
H = T +V(x,t)+ V_{\rm trans}(y,z),
\end{equation}
where $T$ is the kinetic energy, $V(x,t)$ is a slowly varying time-dependent potential
along $x$ and $V_{\rm trans}$ is the time-independent transverse wave-guide potential.
The wave-function can then be written as a product
$\Psi(x,y,z,t)=\Psi_{\rm trans}(y,z)\Psi(x,t)$ where the transverse part $\Psi_{\rm trans}$
is assumed to correspond to the transverse ground-state
at all times, so that we are left with a one-dimensional
problem.
Here, we are not interested in the electric field responsible for the potential $V(x,t)$, but
work with the potential directly.  We choose a rather simple temporal-spatial form,
which will turn out to be sufficient for the control tasks at hand. More complicated
forms could be achieved, but will not be considered in this paper. The potential is
parameterized, and the parameters have to be optimized in such
a way, that the target is achieved. For
optimization we use a simplex-downhill algorithm, with Nelder-Mead
parameters \cite{numrec}, together with a random walk.

At time $t$ = 0 our initial wave packet in the longitudinal coordinate is a
Gaussian,
\begin{equation}
\Psi(x,t=0) = \frac{1}{\sqrt{\sigma\sqrt{\pi}}} e^{-x^2/(2\sigma^2)+ip_0x/\hbar}
\label{eqpsi0}
\end{equation}
where the amplitude $|\Psi(x,t=0)|$ has a full-width at half-maximum (FWHM) of
10 $\mu m$.
The initial momentum has been chosen to correspond to a center-of-mass velocity
$v_{CM}$ of 10 cm/s.
This initial shape of the wave-packet may be generated by starting from a ground-state
in a harmonic trap, which is then released and accelerated to the velocity $v_{CM}$.

For comparison, and in order to estimate the robustness of our optimized external
potentials, we repeat the propagations with the Time-Dependent Gross-Pitaevskii
Equation (TD-GPE), which governs the evolution of BECs with a repulsive
inter-particle contact interaction.
There, $\Psi$ is interpreted as a single particle
function/orbital, which evolves in the mean field of all other atoms.
\begin{equation}
\dot{\Psi} =  -\frac{i}{\hbar} (H + \frac{4\pi\hbar^2\alpha_sN}{m} |\Psi|^2) \Psi
\end{equation}
The mean field potential contains the s-wave scattering length $\alpha_s$,
which for $^{87}$Rb in the $|2,2>$ state is 95.5 $a_0$
\cite{harber}. $N$ is the total
number of atoms. The mean field is repulsive and proportional to the density at
position ${\bf r}$. To reduce the TD-GPE into a one-dimensional form, we assume
that the transverse potential has the harmonic form $V_{\rm trans}(y,z)
=\frac{1}{2}m\omega^2(y^2+z^2)$ such that the transverse ground-state energy
 $\hbar\omega$
of the ground state is larger than the repulsive potential. In this case the TD-GPE
reduces to a 1D form in which the scattering length $\alpha_S$ is replaced with
$\alpha_S\rightarrow \alpha_S/a_{\perp}$, where $a_{\perp}=\pi\hbar/m\omega$ is the
effective area occupied by the ground-state in the transverse directions. For the
sake of better comparison, we use the same initial wave-packet $\Psi(x,t=0)$ as in the TD-SE
case.

\section{Results}

As mentioned in the Introduction, we have chosen four control tasks to study the
versatility of this approach. The following subsections report results for
focusing, acceleration, reflection, and stopping, and discuss the physics
and the robustness of the solutions.

For all four tasks we have chosen an external potential of the form
\begin{equation}
V(x,t) = -V_0\exp\left(-\left(\frac{t-t_0}{\tau}\right)^4\right) \cos^2 \left(
\pi \frac{x-x_c}{\sigma_x}+\phi \right)
\label{extpot}
\end{equation}
 for $|x-x_c|\le \sigma_x/2$, and $V(x,t)=-V_0\sin^2\phi$ otherwise. For the three first 
 tasks we use
$\phi=0$, namely, the potential is attractive at the center $x=x_c$ and forms a
well with a minimum at the center. For the stopping task
we use $\phi=\pi/2$, such that the potential is attractive far from the center
and zero at $x=x_c$, forming a potential barrier.
The times $t_0$ and $\tau$, the width $\sigma_x$, the depth $V_0$ and the potential center
$x_c$ are chosen with constraints, such that the potential is experimentally feasible. The
parameter $x_c$ was chosen to be fixed in the laboratory frame
for the first three tasks and has a time-dependent form $x_c=x_0+vt$ (with two optimization
parameters $x_0$ and $v$) for the stopping task.
We have tested other functional forms for $V(x,t)$, like a Gaussian in time and a
simple half-cosine in space, and found that this does not change the quality or robustness
with which the control tasks are achieved.

\subsection{Focusing}

\begin{figure}[hbt]
\epsfig{file=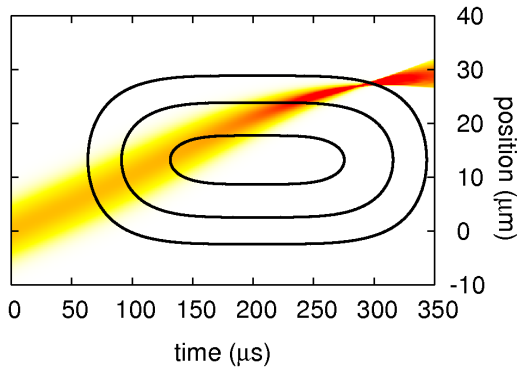,width=7cm}
\caption{Focusing of a wave packet. The time-dependent external potential is given by
contour lines in steps of 30$\mu$K.}
\label{figfocus2b}
\end{figure}

The first target is to focus the pulse, i.e. the external potential should
minimize the width of the wave packet
$\Delta x=\sqrt{\langle x^2 \rangle - \langle x \rangle^2}$. The optimized external
potential is shown in Fig. \ref{figfocus2b} as contour lines (optimal parameters can be
 found in Table \ref{opttab}), together with the time-dependent density from a solution
of the TDSE, i.e. for a  single $^{87}$Rb atom.

At its minimum, $\Delta x_{min}$ =  0.1079 $\mu m$, a factor $F$ = 27.8 smaller
compared to its initial width. This value could have been increased further, if the
external potential had been allowed to be arbitrarily wide. However, to remain within
typical experimental parameters, the width of the potential had been fixed to $\sigma_x$
= 50 $\mu m$ (in the next tasks 50 $\mu m$ was left as an upper limit). Therefore, we
optimized for this goal only the three parameters $V_0$, $t_0$, and $\tau$, as
presented in Table \ref{opttab}. $V_0$ was restricted to be about 100
$\mu$K or less. Such a trap depth has been achieved with an electric field even at the
considerable height of 50 $\mu m$ \cite{schmiedmayer03}.

\begin{table}[hbt]
\begin{tabular}{c|c|c|c|c}
{} & Focus & Acceleration & Reflection & Stopping 
\\ \hline
$V_0$ ($\mu$K)      & 97.73       & 100.0 & 131.4  & 100.0 
\\
$t_0$ ($\mu$s)      & 203.4       & 32.50 & 284.9 & 267.4 
\\
$\tau$ ($\mu$s)     & 134.3       & 35.83 & 136.0 & 103.5 
\\
$\sigma_x$ ($\mu$m) & 50.0        & 34.03 & 31.93 & 52.36 
\\
$x_c$ ($\mu$m) & 13.23        & 13.23 & 13.23 & -0.2927+0.4297$v_{CM}t$ 
\\
$\phi$ & 0 & 0 & 0 & $\pi/2$
\end{tabular}
\caption{Optimal parameters of the external potential for the control
tasks. For focusing the task was to minimize the width $\Delta x$, for acceleration the
kinetic energy was doubled, for reflection the momentum inverted, and for stopping the 
kinetic energy minimized.}
\label{opttab}
\end{table}

The physics of the focusing process is rather simple:
In this example the external potential is switched on when the wave packet is over a region with
a relatively small gradient and as time evolves the atoms are facing a hill. Focusing is
explained by the gradient of the force, so that a different force is applied by the
potential on the different parts of the wave-packet. The parabolic shape of the potential
implies that while the wave-packet is climbing up the potential slope at $x>x_c$ the
front part of the wave-packet experiences a stronger retarding force in the backward
direction relative to the trailing part. This causes these two parts to run into each
other while the wave-packet becomes narrower until the front and rear parts reach each
other at maximum focusing and then pass each other and continue to move further away as
the wave-packet expands.
In the case of a BEC, the interaction between the particles may modify this
dynamics, as we show below.

\begin{table}[hbt]
\begin{tabular}{c|c|c|c|c}
{} & Focus & Acceleration & Reflection 
\\ \hline
$V_0$ &      0.147 & 0.275 & 0.523 
\\
$t_0$ &      0.674 & 0.130 & 0.152 
\\
$\tau$ &     0.277 & 0.181 & 0.857 
\\
$\sigma_x$ &   -   & 0.200 & 0.920 
\end{tabular}
\caption{Stability of target achievement, in percent. No stability analysis was done
for the stopping task (see text).}
\label{stabtab}
\end{table}

How sensitive is the focus factor $F$ to errors in the parameters of the
external potential? In order to answer this, we changed each of the optimal
parameters by $\pm$1 \%, and re-ran the simulations. Table \ref{stabtab} reports
the mean of the modulus of the change in the minimal width, given in percent of the
optimal $\Delta x_{min}$ reported above. For all three parameters, the change is
less than 1 \%, which implies good stability against external perturbations.

\begin{figure}[hbt]
\epsfig{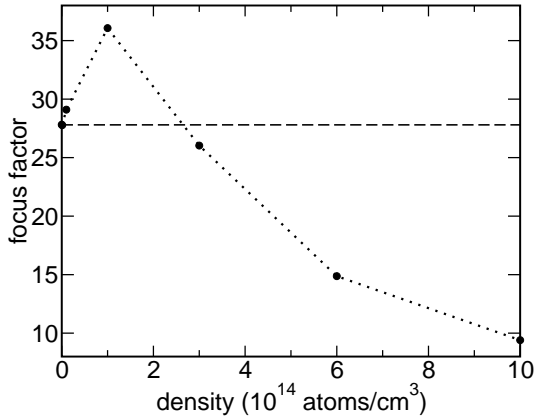}
\caption{Focusing efficiency as a function of BEC density. The dashed horizontal
line indicates the TD-SE solution.}
\label{focusdens}
\end{figure}

\begin{figure}
\epsfig{file=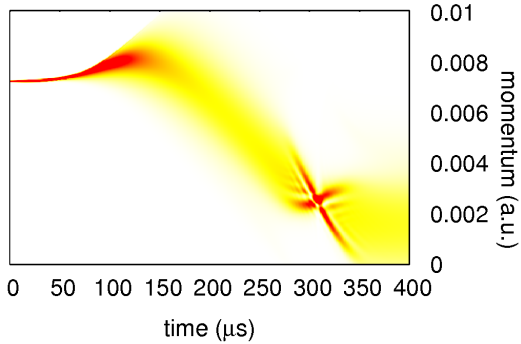,width=7cm}
\caption{Time-dependent momentum density of a BEC that is focused in position space.
At the time when the spatial width is minimal, an additional feature appears, which
is not present in the Schr\"odinger case.}
\label{figtdgpemom}
\end{figure}

At this point the question arises, how a BEC would behave. For the same set of
optimized parameters and the same initial $\Psi$, that were used
for the TD-SE, we solved the TD-GPE for various densities. Roughly, the time evolution
looks like the one in Fig. \ref{figfocus2b}. The width of the BEC is slightly larger at
early times, because of the repulsive potential, but this could not be discerned given
the scale of Fig. \ref{figfocus2b}.  The expectation then is, that the repulsive mean
field potential counteracts the focusing. However, we find that this is not
strictly true, see Fig. \ref{focusdens}. For a certain range of densities, the
focusing is even more efficient. This can be best understood by considering the motion
of the frontal and trailing parts of the wave-packet in the center-of-mass frame of
reference. Unlike the case of non-interacting particles, where these parts of the
wave-packet can freely pass through each other, in the case of a BEC the repulsive
interaction potential counteracts the movement of the trailing and frontal parts
through the central part when they are already close together. As the shape of the
wave-packet is not Gaussian at this stage and contains significant tails, this
counter-action serves to concentrate these parts better around the center. Another
implication of this process is that the momentum distribution narrows down significantly
near the maximum focusing points, as shown in Fig.~\ref{figtdgpemom}. This focusing
both in position and momentum space appears only in the case of the BEC and is absent
when the TD-SE is solved as no significant change in the momentum distribution is seen.
If the density of the BEC is increased further, then the expected decrease of the
focusing efficiency is observed.

\begin{figure}
\epsfig{file=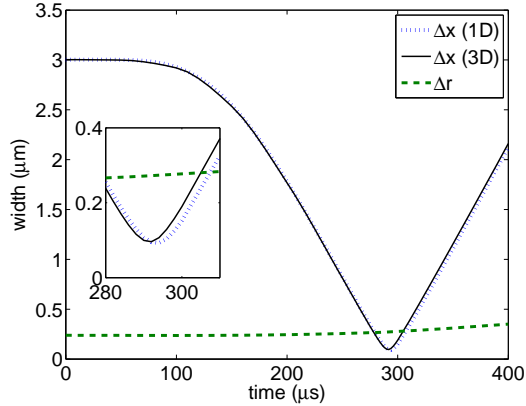,width=7cm}
\caption{Mean longitudinal and transverse width of a BEC wave-packet during the
focusing process. The longitudinal focusing efficiency is almost similar to that
obtained in a 1D simulation, and the transverse expansion is not significant
while the maximum longitudinal focusing is reached.}
\label{BEC_cyl}
\end{figure}

A question that may arise is the validity of the one-dimensional approximation for the
dynamics of a BEC when it is focused. It may appear that even if initially the repulsive
energy is smaller than the transverse confinement potential and the transverse part of
the wave function is roughly the ground state of the transverse potential, then after
focusing the density is increased significantly and the wave-packet may experience
significant expansion in the transverse direction. In order to resolve this we have
solved the TD-GPE in cylindrical coordinates for the case of a BEC with $N=1000$
atoms and transverse confinement of $\omega_{\rm trans}=2\pi\times 400$Hz
(initial peak density of $1.45\cdot 10^{14}$ atoms/cm$^3$).
As shown in Fig.~\ref{BEC_cyl}, the focusing efficiency is almost unchanged relative
to the one-dimensional approximation and the transverse expansion is not very
significant before and at the time of maximum longitudinal focusing. The effect of
the transverse repulsion leads to significant expansion in the transverse direction
only after maximum longitudinal focusing is reached, when the BEC significantly expands
in all directions.

\subsection{Acceleration}

\begin{figure}[hbt]
\epsfig{file=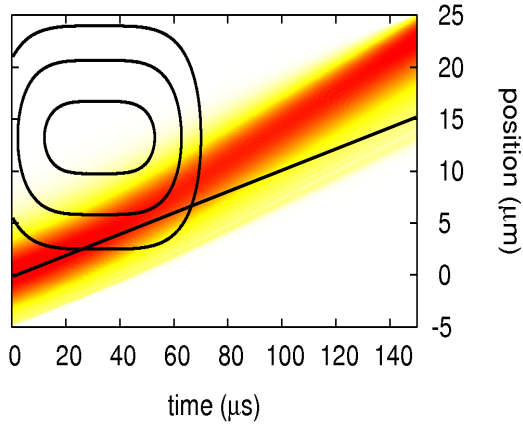,width=7cm}
\caption{Accelerated wave packet. The straight line indicates the center of mass
without acceleration. The time-dependent external potential is given by
contour lines in steps of 30$\mu$K.}
\label{sqrt2pfig}
\end{figure}

The second control task is to increase the momentum of the wave packet. As an example
of a possible application, one may imagine a train of atom laser pulses, where an
acceleration protocol can be a way to adjust the separation of the packets. In another
application it might be used to change the energy or de-Broglie wave length. The target
for our second optimization has therefore been chosen to increase the kinetic energy by
a factor of 2. The main constraint is the limit that we have put on the maximal depth
of the potential (of about 100 $\mu$K). This limitation may be overcome by performing
the task repeatedly, using a sequence of potential wells.
For example, in the 'downhill' acceleration described here,
once the wave packet reaches the end of the potential, a new potential would appear and
the acceleration process would continue. For simplicity we will not discuss these
extensions in the framework of this paper.

Fig. \ref{sqrt2pfig} shows the time-dependent density and potential. The straight black
line projects the position of the center of mass if momentum did not change. When the
wave packet enters the region of the attractive potential, it gains energy, until the
potential is shut off at the right time. The final momentum changes by less than
0.3 \%, if the parameters of the potential are changed by $\pm$ 1\%, see table
\ref{stabtab}. The parameter $t_0$ = 32.5 $\mu s$ is the shortest timescale involved
in this study. It should be noted that a precision of 1\% here means to switch the
electrodes with a precision of about 3 MHz or better, which is typically doable. A
simulation with the TD-GPE, using a density of $10^{15}$ atoms/cm$^3$, gives a result
which is almost identical to the TD-SE: The final momentum exceeds the target momentum
by a mere 0.08 \%.

\subsection{Reflection}

\begin{figure}[hbt]
\epsfig{file=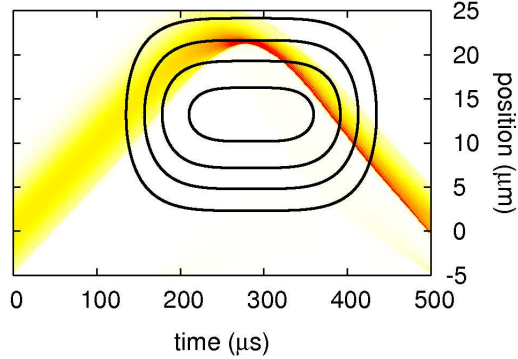,width=7cm}
\caption{Reflection of a wave packet. The time-dependent external potential is given by
contour lines in steps of 30$\mu$K.}
\label{reflectfig}
\end{figure}

The third goal is to reflect the wave packet, i.e. to invert the momentum. This
kind of coherent mirror is an essential part of matter wave optics. The
first solutions found by the optimization algorithm were based on quantum
reflection, but these were very unstable against perturbations in the
parameters. In order to obtain a good result, we had to allow a slightly larger
potential well depth, see Table \ref{opttab}. The solution is shown in Fig.
\ref{reflectfig}.

At around $t$=150 $\mu s$, when the wave packet is over the
center of the potential, it is switched on. Because at that time and position
the wave packet does not feel a gradient of the potential, it is not
accelerated, while the potential is lowered with time. At around $t$=200 $\mu s$ the
wave packet hits  an almost time independent potential wall. At around 270
$\mu s$ all its kinetic energy has been transformed into potential energy, and it
is reflected. The final momentum was about 0.1 \% smaller than $p_0$, due to the
influence of the time dependent potential, which also perturbs the initial Gaussian shape
 of the wave packet. Again we found good stability, see Table \ref{stabtab}. The
 final momentum from the solution of the TD-GPE deviated by less than 0.01 \%.

\subsection{Stopping}

\begin{figure}
\epsfig{file=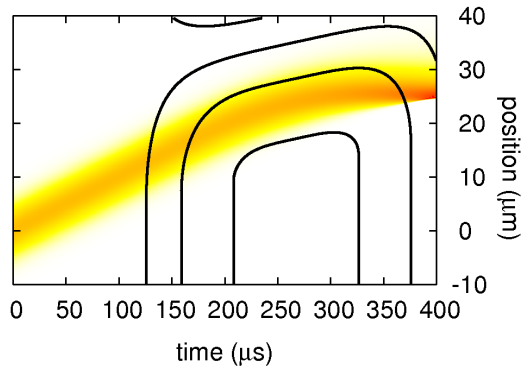,width=7cm}
\caption{Stopping of a wave packet at a mobile barrier, created by adiabatically
lowering the external potential. The contour lines of the potential are at 3$\mu$K,
30$\mu$K, and 90$\mu$K.}
\label{stopdens}
\end{figure}

\begin{table}[hbt]
\begin{tabular}{c|c|c|c|c|c}
{} &  $t_0$  &  $\tau$  &  $\sigma_x$  &  $x_c$ offset  &  $x_c$ slope \\ \hline
+1\% & 0.909 \%&  0.912 \% &  0.932 \% &  0.903 \% &  0.919 \% \\
-1\% & 0.909 \%&  0.910 \% &  0.898 \% &  0.902 \% &  0.890 \%
\end{tabular}
\caption{Remaining kinetic energy in percent of the initial kinetic energy, when the
parameters of the stopping potential are changed by $\pm$1 \%. The original set of
parameters led to 0.903 \% remaining.}
\label{sensstop}
\end{table}

In this example our goal is to bring the atomic wave-packet to an almost complete
stop, not just by achieving a zero final momentum of the center of mass, but also by
minimizing the total kinetic energy of the target wave-packet.
Different from the previous tasks, here the central
point $x_c$ of the potential is not fixed in time.
In classical mechanics, a particle is stopped when it collides with a much heavier
particle moving in the same direction at half the velocity. For the present case,
this means a collision with a mobile potential barrier, which is created from the
purely attractive external potential of Eq. (\ref{elecpot}),
by lowering the potential in front of the barrier. To describe this situation,
we use the same form of the potential of Eq.~(\ref{extpot}) with $\phi=\pi/2$,
representing a potential barrier with a moving center $x_c=x_0+vt$, which is more
attractive far from this center.
For this task, we have chosen $V_0$ to be fixed at 100 $\mu$K, so that the potential
has a total of five parameters to be optimized. A perfect stopping implies a kinetic
energy of exactly zero, but this can only be achieved for an infinitely wide wave
packet. Therefore, we limited the optimization procedure to a reduction of the kinetic
energy to 0.903 \%, see Fig. \ref{stopdens}. As this is not an extremum of the kinetic 
energy, stability is not achieved in the normal sense. However, we give the 
reduction of the kinetic energy for parameters changed by $\pm$1 \% in table 
\ref{sensstop}, from which the sensitivity of the mechanism can be estimated. 
A comparison with the dynamics of a BEC gave a very similar result, with a 
final kinetic energy which was lower by 0.2$\mu$K than that achieved for 
non-interacting particles.

\section{Splitting}

\subsection{Splitting into two parts}

\begin{figure}
\epsfig{file=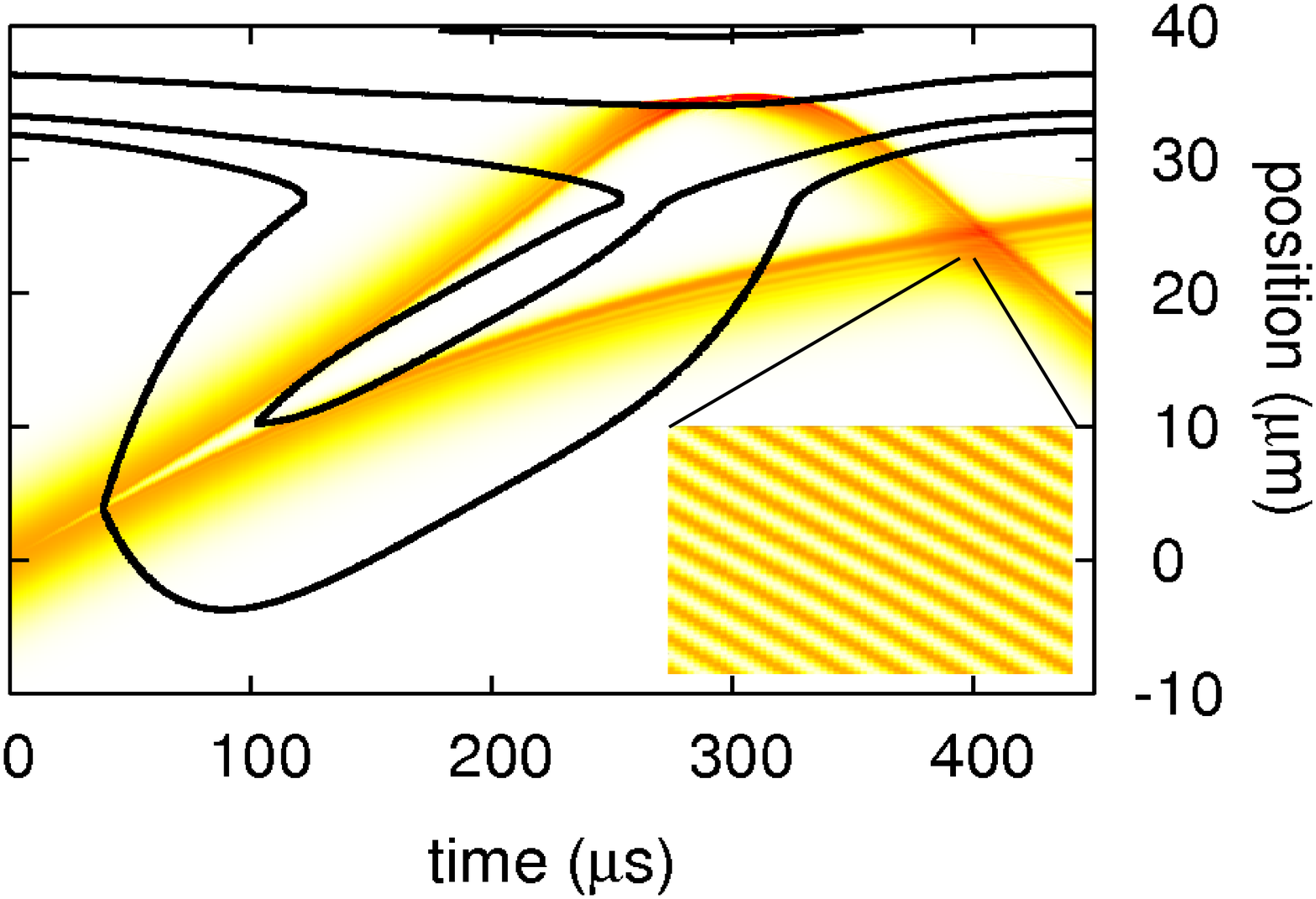,width=7cm}
\caption{Splitting of the wave packet. The time-dependent external potential is
given by contour lines at 60$\mu$K, 90$\mu$K, 180$\mu$K, and 360$\mu$K.
Inset: a closer look at the intersection region, revealing the interference pattern.}
\label{figsplit}
\end{figure}

In order to show that this method may be used also for complex tasks,
we turn our attention to the goal of coherently splitting the wave packet.
We have chosen a double Gaussian target density, separated by 4$\sigma$ of the
initial width of the wave packet, see Eq. (\ref{eqpsi0}). The absolute position,
where the target is achieved, was not fixed. 

For the splitting we employed a potential
\begin{equation}
V(x,t) = -2V_0\exp\left(-\left(\frac{t-t_0}{\tau}\right)^4\right)
\frac{|x-v_{CM}t|}{\sigma_x} ,
\end{equation}
for $|x-V_{CM}t|\le \sigma_x/2$ and $V(x,t)=-V_0$ otherwise.
This is a potential barrier that travels with the initial center-of-mass velocity
mentioned above, and
has a total width of $\sigma_x$.
\begin{figure}
\epsfig{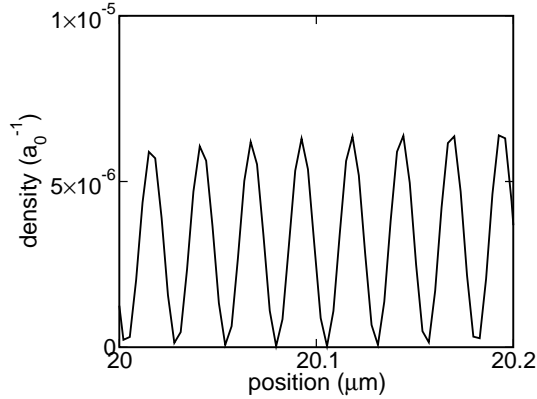}
\caption{Interference pattern during the crossing of the two partial beams.}
\label{figfinden}
\end{figure}

Fig. \ref{figsplit} shows the density as a function of time, for parameters $V_0$=
94.6$\mu$K, $t_0$=189.0 $\mu$s, $\tau$=183.4 $\mu$s, and $\sigma_x$=
82.2 $\mu$m. Because a maximum of the potential energy is created under the center of
the moving wave packet, it splits neatly into two parts. In addition, we added a
parabolic barrier at large $x$, so that one part is reflected and can interfere with
the second partial beam. When the two cross, at approximately t=405 $\mu$s, an
interference pattern appears, which depends on the relative momenta of the two parts.
Fig. \ref{figfinden} shows in detail, at a specific time, the atom density during the
crossing. As expected, the fringe visibility is practically 100$\%$ as no dephasing
mechanisms have been introduced. These are beyond the scope of this paper.

Last, we again repeated the splitting calculation with the TD-GPE. We assumed a
relatively high density of 10$^{15}$ atom/cm$^3$, and no noticeable differences have
been found, compared to the TD-SE case.

\subsection{Splitting into three parts}

Finally, we use the most complex task yet, the splitting into three peaks, to emphasize 
the uniqueness of our scheme relative to previous schemes 
\cite{Castin96,Japha02,Ott03,Oberthaler03,Oberthaler04,Cornell05}. In the introduction 
we noted that our new scheme aims at making a step forward in two ways: first, optical 
schemes have the disadvantage of not being feasible very close to a surface because of 
light diffraction. This limits the potential usefulness of such schemes to being applied 
to surface chemistry. Second, all the previous schemes utilize potentials with a high degree 
of symmetry either in the lab frame or in some moving frame. This symmetry makes the task 
of creating arbitrary wave forms, which are in general asymmetric, hard to realize. In 
order to demonstrate this, we present in the following the creation of an asymmetric wave form.

\begin{figure}
\epsfig{file=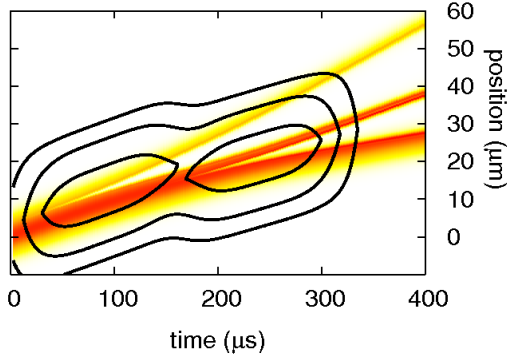,width=7cm}
\caption{Asymmetric splitting of the wave packet into three parts. Here we clearly show that 
the time dependent potentials suggested here may take an asymmetric form, enabling the creation 
of asymmetric matter-wave pulses.}
\label{figsplit3pot}
\end{figure}

\begin{figure}
\epsfig{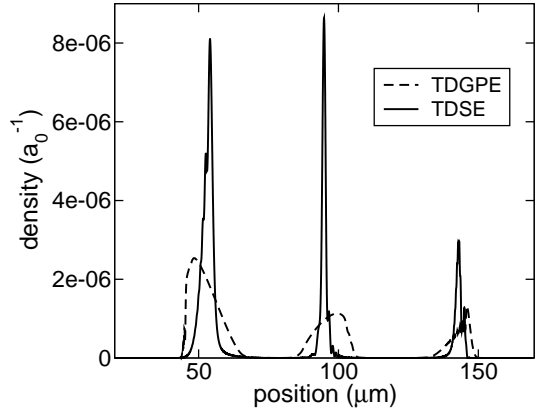}
\caption{Asymmetric splitting of the wave packet into three parts. The target ratio
for the integrated density of the three peaks was 4:2:1. The solid line shows the
result from the TD-SE, the dashed line from the TD-GPE. The latter shows broader and
smoother peaks, because of the repulsive atom-atom interaction.}
\label{figsplit3}
\end{figure}

We have chosen to split the wave packet asymmetrically into three unequal parts, so 
as to form an asymmetric wave form. The same functional form of the external 
potential as in the previous subsection has been chosen, but it was applied 
twice, see Fig. \ref{figsplit3pot}. The target ratio of integrated densities 
under the three peaks was chosen to be 4:2:1. This was achieved with very good 
accuracy: We obtained 3.999:2.003:0.999. When the same potential was used to 
split a BEC, this deteriorated only slightly to 3.993:2.001:1.006, see Fig. 
\ref{figsplit3}. Although the target was achieved with almost the same 
quality, there is a big difference in the actual shape of the peaks, as the 
repulsive force between the atoms smoothes and broadens them significantly.

Let us briefly describe the details of this task. The total potential $V(x,t)$ 
is given by the maximum of
\begin{equation}
V_0e^{-((t-t_0)/\tau)^4}\left( 1-\frac{|x-(v_{CM}t+x_1)|}{\sigma}\right)
+V_0e^{-((t-(t_0+2\tau))/\tau)^4}\left( 1-\frac{|x-(v_2t+x_2)|}{\sigma}\right)
\end{equation}
and zero. Only the velocity $v_2$ and the offsets $x_1, x_2$ were
optimized. $V_0$,$t_0$,$\tau$ and $\sigma$, were kept constant with values 40 
$\mu$K, 90 $\mu$s, 75 $\mu$s and 20 $\mu$m, respectively.

The target functional was constructed as follows: If the integrated norm under
peak $i$ is $N_i$ then
\begin{displaymath}
\mbox{Deviation} = \sqrt{(N_1-4/7)^2+(N_2-2/7)^2+(N_3-1/7)^2}
\end{displaymath}
measures the deviation from the target (evaluated at $t=400 \mu$s). The best
parameter set ($x_1$=3.207 $\mu$m, $v_2$=8.000 cm/s, $x_2$=1.778 $\mu$m)
achieved a deviation of 0.000231. If $x_1(x_2)$ is changed by $\pm$1 \%, the 
average deviation was found to be 0.00341(0.00328), which is equivalent to 
about 1 \% in the integrated norm under the peaks. We note that a 1 \% change 
in position is equivalent to the few tens of nano-meter fabrication resolution 
we expect in the electrode array producing the potential. Last, if $v_2$ is 
changed by $\pm$1 \%, the average deviation is 0.02912.

\section{Summary and Outlook}

We have shown that considerable control over the shape of an atomic wave-packet is
possible by using a quasi-static potential. As a specific realization one may utilize
the weak electric field from an array of electrodes on an atom chip.

We have shown that four basic control tasks, namely, focusing, acceleration, reflection, 
and stopping, can be performed. The control schemes were found to be robust, both with
regards to the parameters of the external potential, and the density of the BEC. The 
non-trivial dependence of the focusing efficiency on the density of the BEC, and the 
narrowing of the momentum distribution may also prove to be useful tools in future 
applications. We have also described the application of our method to coherent splitting.

Finally, we have described the use of an asymmetric potential for the creation of an 
asymmetric wave form. Such versatility is the main advantage of our scheme.

We have limited ourselves in this study to a rather simple external potential, while an
array of electrodes is capable of much more complex forms. The limits of this remain to
be investigated in the future. Ultimately, the shaping of atom laser wave packets is
not a goal in itself, but rather a tool to achieve control in other areas, such as
ultra cold chemistry.

\section{Acknowledgements}
This work was supported by the The German-Israeli Binational Science Foundation (GIF), 
and the Israeli Science Foundation.

\end{document}